\documentclass[]{JHEP3}
\usepackage{amsfonts} 
\usepackage{amsbsy} 
\usepackage{amssymb} 
\usepackage{epsfig}
\usepackage{psfrag}
\usepackage{cite}

\newcommand{\B}[1]{{\mathbb #1}}
\newcommand{\C}[1]{{\mathcal #1}}

\newcommand{\beq}{\begin{equation}}
\newcommand{\eeq}{\end{equation}}
\newcommand{\bea}{\begin{eqnarray}}
\newcommand{\eea}{\end{eqnarray}}
\newcommand{\rf}[1]{(\ref{#1})} 

\newcommand{\Tr}{{\hbox{Tr}\,}}

\newcommand{\absval}[1]{{\left\vert#1\right\vert}}
\newcommand{\norm}[1]{{\left\vert \left\vert #1 \right\vert \right\vert}}

\newcommand{\half}{{1\over 2}}

\newcommand{\pa}{\partial}

\newcommand{\zbar}{{\overline{z}}}

\newcommand{\vep}{\epsilon} 
\newcommand{\kt}{\rangle}
\newcommand{\br}{\langle}

\title{Scalar Solitons on the Fuzzy Sphere}
\author{Peter Austing\\Science Institute, University of Iceland, Dunhaga 3,
107 Reykjavik, Iceland\\E-mail: \email{austing@raunvis.hi.is}}
\author{Thordur Jonsson\\Science Institute, University of Iceland, Dunhaga 3,
107 Reykjavik, Iceland\\E-mail: \email{thjons@raunvis.hi.is}}
\author{Larus Thorlacius\\Science Institute, University of Iceland, Dunhaga 3,
107 Reykjavik, Iceland\\E-mail: \email{lth@raunvis.hi.is}}

\abstract{We study scalar solitons on the fuzzy sphere at 
arbitrary radius and noncommutativity.  We prove that no solitons exist 
if the radius is below a certain value.  Solitons do exist for radii 
above a critical value which depends on the noncommutativity parameter.  
We construct a family of soliton solutions which are  
stable and which converge to solitons on the Moyal plane in an 
appropriate limit.  These solutions are rotationally symmetric about
an axis and have no allowed deformations. Solitons that describe multiple lumps on the fuzzy sphere can also be constructed but they are not stable.}

\keywords{Solitons Monopoles and Instantons, Non-Commutative Geometry}

\preprint{RH-10-2002}
\begin{document}

 \section{Introduction}  
It is an interesting fact that the classical equations of motion of
noncommutative field theory have a richer set of solutions than their 
commutative counterparts. Noncommutative solitons include deformations of conventional solitons but also new objects that are absent in the commutative theory. This applies to both gauge theories and scalar field theories and can be traced to the existence of a new length scale associated with the noncommutativity. A striking example is provided by scalar theory on the Moyal plane \cite{Gopakumar:2000zd}, where the corresponding commutative theory can have no solitons by Derrick's theorem \cite{Derrick:1964ww}. Several authors have found explicit soliton solutions in gauge theory with and without matter fields, see, e.g., 
\cite{Polychronakos:2000zm,Aganagic:2000mh,Gross:2000ss,Harvey:2000jb}
and noncommutative deformations of commutative solitons have been studied
in \cite{Jatkar:2000ei, Bak:2000ac, Lozano:2000qf}. For reviews of noncommutative quantum
field theory we refer to \cite{Szabo:2001kg,Harvey:2001yn}.

The scalar solitons on the Moyal plane were first observed in the limit of infinite noncommutativity where the kinetic term in the action can be neglected and solutions correspond to multiples of projectors \cite{Gopakumar:2000zd}. This line of reasoning was generalised to scalar fields on arbitrary K\"{a}hler manifolds including the
sphere in \cite{Spradlin:2001ku} and subsequently multi-solitons on the fuzzy sphere
in the infinite noncommutativity limit were considered in \cite{Vaidya:2001rf}.
In \cite{Durhuus:2000uz,Durhuus:2001nj}  it was shown that there exist scalar solitons on the Moyal plane at large but finite
values of the noncommutativity parameter and that no soliton solutions 
exist depending smoothly on small values of the noncommutativity parameter as expected from 
Derrick's theorem. All the solitons at
finite noncommutativity constructed so far have the property that they are
rotationally invariant about some point.  It is not known whether there
exist stable soliton solutions in the plane describing separated lumps, but
there are good reasons to believe that no such solutions exist and
that all soliton solutions in the plane are rotationally invariant about some 
point.  First of all, separated lumps attract each other according to 
perturbative  calculations at large noncommutativity parameter 
\cite{Gopakumar:2001yw,Lindstrom:2000kh}.  Furthermore, it has recently been 
proven that there cannot exist a family of static solutions on the 
Moyal plane at finite noncommutativity that interpolates smoothly between the 
solution describing two overlapping solitons and a solution with two 
infinitely separated solitons \cite{Durhuus:2002fh}.  Finally, the stability
results for scalar solitons on the fuzzy sphere obtained below, combined 
with a scaling limit that yields the Moyal plane from the fuzzy sphere
(see {\it e.g.} \cite{Chu:2001xi}), provide further evidence for the 
non-existence of stable multi-lump solutions. 
Further results for noncommutative theory in the plane
can be found in \cite{Araki:2001hu,Hadasz:2001cn,Solovyov:2000xy,Matsuo:2000pj,Jackson:2001iy,Zhou:2000xg}, and on the fuzzy sphere \cite{Hikida:2000cp,Hashimoto:2001xy,Steinacker:2000yk,Kimura:2001uk,Chan:2001yi,Dolan:2001gn,Iso:2001mg,Dolan:2002an}. 

In the present paper, we study the equations of motion for a scalar field on 
the fuzzy sphere in detail, paying particular attention to issues that arise 
at finite noncommutativity. Part of our analysis is parallel to that of 
\cite{Durhuus:2000uz,Durhuus:2001nj} which dealt with existence and stability 
of scalar solitons on the Moyal plane. The algebra underlying 
the fuzzy sphere is finite dimensional, and this simplifies the mathematics, 
especially the stability analysis.  It enables us to establish some uniqueness 
results by ruling out deformations of the field profile of a stable soliton. 
We are also able to prove a stronger nonexistence result at 
small noncommutativity than is available on the Moyal plane. 

The fuzzy sphere is characterised by two parameters: $R$ which can be interpreted as the radius of the sphere, and $j$ which labels the associated matrix algebra. Together these parameters determine the strength of noncommutativity. A scalar field theory on the fuzzy sphere is further characterised by a scalar field potential $V(\phi)$ which we take to be a double well potential with a global minimum at the origin. Our main results are then the following:

\begin{itemize}

\item {\it There are no solitons for small values of the radius.}

\item {\it As the radius is increased a large number of soliton solutions
come into existence, most of which are unstable.}

\item {\it A necessary condition for a soliton, that is rotationally invariant 
around the north-south axis, to be stable is that its eigenvalues form a 
decreasing sequence (in the standard $su(2)$ basis). At sufficiently large 
radius, another condition is that the eigenvalues lie close to the minima of 
the potential, and these two conditions are also sufficient.}

\item {\it The only small deformations of a stable soliton
which lead to new soliton solutions are rotations.}

\item {\it There is a family of stable, rotationally invariant solitons on the 
fuzzy sphere which converge to stable solitons in the Moyal plane as both $R$ 
and $j$ tend to infinity keeping $R^2/j$ fixed and sufficiently large.}

\end{itemize}

The limit of infinite noncommutativity is by no means smooth, and the set of 
classical solutions is drastically reduced at finite as compared to infinite 
noncommutativity. The difference is particularly striking when one considers 
multisolitons. In fact, the very notion of a multisoliton solution is somewhat 
problematic at finite noncommutativity since there is no precise definition 
of soliton number in this case. In the limit of infinite noncommutativity, 
the soliton number of a stable classical solution can be taken to be the rank 
of the corresponding projector, but solutions at finite noncommutativity are 
no longer given by projectors and this definition fails. A physicist might 
wish to define a multisoliton as a configuration with two or more localised 
lumps in the scalar field that are separated on the fuzzy sphere, but this is
not a precise mathematical definition. Our results strongly suggest that there are in fact no stable solutions with 
separated lumps, and that all stable solutions are rotationally symmetric. 
There is, in other words, no moduli 
space of scalar multisolitons on the fuzzy sphere at finite noncommutativity.
We do, however, give evidence for multi-soliton solutions on the fuzzy sphere 
describing separated lumps that are evenly distributed along a great circle.
These configurations are unstable due to the mutual attraction between lumps. 
In the special case of two identical lumps at the north and south poles, we 
give rigorous arguments for the existence of the unstable solutions. 

By taking an appropriate limit $j,R \rightarrow \infty$, one can blow up 
the region close to the south pole into the Moyal plane.  Assuming that 
every solution in the Moyal plane descends from solutions on the fuzzy 
sphere in the limit, one would expect a solution with separated lumps on 
the Moyal plane to descend from an infinitesimal deformation that breaks 
the rotation symmetry of a solution on the fuzzy sphere. The fact that no 
such deformations exist strongly suggests that there are no stable 
multi-lump solutions at finite theta on the Moyal plane, in agreement 
with the results of \cite{Durhuus:2002fh} and expectations from 
perturbative calculations in the limit of infinite 
non-commutativity \cite{Gopakumar:2001yw,Lindstrom:2000kh}.

The unstable great circle solitons are lost in the Moyal plane limit since 
any lump not sitting at the south pole will be moved to infinity.  By symmetry 
we would, however, expect the existence of an unstable solution consisting of 
an infinite chain of lumps finitely separated in a straight line on the Moyal 
plane. Such a solution would be periodic in one direction and have infinite 
energy.  It could presumably be further generalized to a regular 
two-dimensional array of lumps.

The paper is organized as follows.  In the following section we give a brief
introduction to the fuzzy sphere and establish our notation.  
In section~\ref{secnon} we prove the nonexistence of static soliton solutions 
at small radius. In section~\ref{seccrit} we study unstable solitons which 
bifurcate from the trivial soliton as the radius is increased. We construct 
some rotationally symmetric solitons on the fuzzy sphere in 
section~\ref{largetheta}, show that they are stable in section~\ref{sec:stab}, 
and that they converge to solitons on the Moyal plane in an appropriate limit 
in section~\ref{planar}.  As a byproduct of the stability analysis in 
section~\ref{sec:stab}, we show that the only possible deformations of these 
stable solitons correspond to rotations that do not affect the shape of the 
field profile. In section~\ref{secunstable} we establish the existence of 
(unstable) solutions that correspond to separate lumps at the north and south 
poles and give arguments for their generalization to lumps evenly distributed 
along a great circle. 

\section{The Fuzzy Sphere} 
\label{secfuzsphere}

One of the simplest interesting noncommutative spaces is the fuzzy sphere
\cite{Madore:1992bw} which can be regarded as an approximation to the true sphere
obtained by cutting off the function algebra on the sphere by requiring the
total angular momentum quantum number to be smaller than a prescribed value
$j$.  This requires of course that the ordinary product of functions be
modified to a noncommutative product.  If $x_1$, $x_2$ and $x_3$ are
Cartesian coordinates in $\B R^3$ and the sphere is defined as
\beq\label{0}
x_1^2+x_2^2+x_3^2=R^2,
\eeq
then the commutation relations between the $x_i$'s become
\beq\label{00}
[x_j,x_k]=i\sigma\vep_{jk\ell}x_\ell.
\eeq
Here $\sigma$ is a positive real parameter.
For our purposes it is convenient to regard the fuzzy sphere as the algebra
$M_j$ of
$(2j+1)\times (2j+1)$ matrices, where $j$ is a half integer.  These matrices
act on a $(2j+1)$-dimensional vector space which we denote by $W_j$.  In $M_j$
we have the spin matrices ($su(2)$  generators) for spin $j$ which
satisfy the usual commutation relation
\beq\label{1}
[J_j,J_k]=i\vep_{jk\ell}J_\ell,
\eeq
and $J_iJ_i=j(j+1)$.
These matrices can be seen to generate the full matrix algebra $M_j$.
With the above conventions we see that the $x_i$ are proportional to the 
angular momentum generators, $x_i = \sigma J_i$, and the noncommutativity 
parameter $\sigma$ is governed by the spin $j$ and and the radius of the 
sphere through 
\beq\label{ncparameter}
R^2=j(j+1)\sigma^2.
\eeq
The commutative sphere of radius $R$ is recovered in the limit 
$j\rightarrow\infty$, keeping $R$ fixed.  Alternatively, one can take
a scaling limit $j,R\rightarrow\infty$, keeping the ratio
$R^2/j$ fixed, and consider a small neighborhood around the south pole.
The fuzzy sphere commutation relations (\ref{00}) can then be shown to
reduce to those of the Moyal plane \cite{Chu:2001xi}.

Scalar field theory on the fuzzy sphere \cite{Grosse:1996ar} 
is defined by the action functional
\beq\label{action}
S= {4\pi \over 2j+1} \Tr \left( [J_i,\phi][\phi,J_i] +  R^2 V[\phi] \right)
\eeq
where $V$ is the potential and $\phi$, the field, is an arbitrary hermitian
$(2j+1)\times (2j+1)$ matrix.  The action is invariant under rotations $\phi \rightarrow \C R \phi \C R^{-1}$ where $\C R= \exp(i \theta\hat{n}_i J_i)$. The variational equation of $S$ is
\beq\label{eqmotn}\label{eqm}
2[J_i,[J_i,\phi]] +  R^2 V'(\phi ) =0.
\eeq
It follows immediately that any solution to this equation satisfies
the condition
\beq\label{trace}
\Tr V'(\phi )=0.
\eeq
Introducing the raising and lowering operators $J_\pm=J_1\pm iJ_2$, the
variational equation can be written
\beq\label{eqm2}
[J_-,[J_+,\phi ]]+[J_3,\phi]+[J_3,[J_3,\phi]]+ { R^2 \over 2} V'(\phi ) =0.
\eeq
Let $|m\kt $ denote the standard basis for $W_j$ consisting of the eigenvectors
of $J_3$, $J_3|m\kt =m|m\kt$, $m=-j, -j+1, \ldots , j$.  If a solution
$\phi$ to Eq.\ \rf{eqm2} is diagonal with respect to this basis, its matrix
elements $\phi_m=\br m|\phi |m\kt$ satisfy the equation
\beq\label{eqmd}\label{diffeq}
\alpha_m(\phi_{m+1}-\phi_m)-\alpha_{m-1}(\phi_m-\phi_{m-1})
= { R^2 \over 2} V'(\phi_m),
\eeq
for $m=-j,-j+1,\ldots ,j$, where $\alpha_m=j(j+1)-m(m+1)$.  It is convenient
to formally introduce $\phi_{j+1}$ and $\phi_{-j-1}$ with the convention 
that $\phi_{j+1}=\phi_j$ and $\phi_{-j-1}=\phi_{-j}$.
Summing eq.\ \rf{eqmd} over $m$ the left hand side telescopes and we obtain
the first order difference equation
\beq
\label{eq1}
\phi_{m+1}-\phi_m={ R^2\over 2 \alpha_m} 
\sum_{i=-j}^m V^\prime (\phi_i) \, , \;\;\; m=-j,\cdots , j-1
\eeq
with the constraint
\beq
\label{eq2}
\sum_{i=-j}^j V^\prime (\phi_i) =0.
\eeq
Diagonal matrices commute with $J_3$ and diagonal solutions of the scalar
field equations correspond precisely to solitons which are invariant under 
rotations about the $z$-axis.

We shall restrict ourselves to studying potentials $V(x)$ which are twice
continuously differentiable, nonnegative, and having a
double zero at $x=0$ and $V(x)>0$ for $x\neq 0$. We shall assume
$V(x)$ has a local minimum (false vacuum) at $s>0$ in addition 
to the global minimum (true vacuum) at $x=0$, that is $V'(x)$ 
has zeros at $x=0$, $x=r$ and $x=s$ 
with $0 < r<s$, and no others (see figure \ref{fig1}).

\FIGURE[thb]{
    \psfrag{V'(x)}{$V'(x)$}
   \psfrag{x}{$x$} 
   \psfrag{0}{$0$} \psfrag{u}{$u$} \psfrag{v}{$v$}
  \psfrag{r}{$r$}
 \psfrag{s}{$s$}
    \includegraphics[width=10cm]{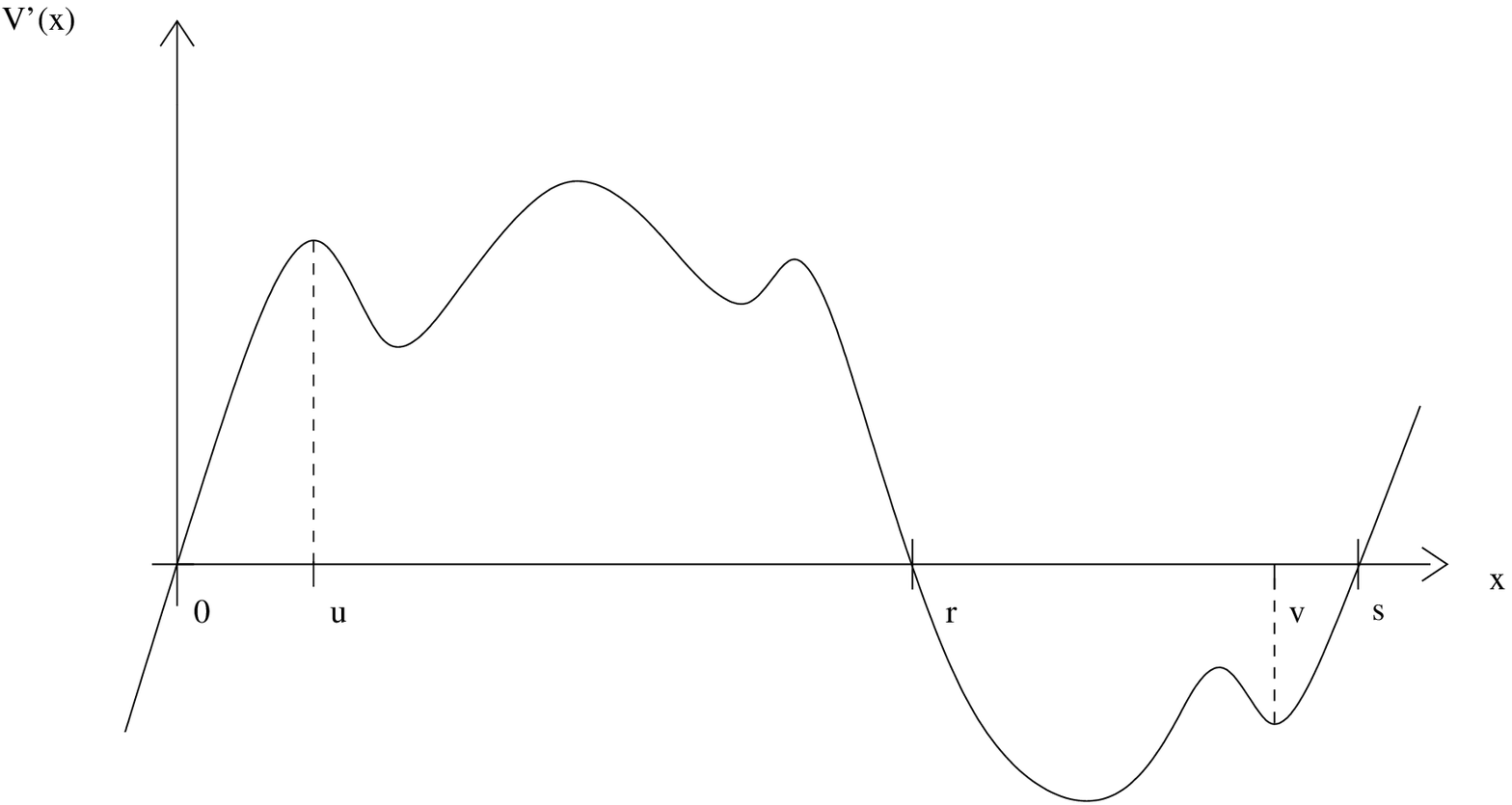}
    \caption{An example of the derivative of a  double-well potential satisfying our assumptions.
    The points $u$ and $v$ are the critical
points of $V'$ closest to $0$ and $s$, respectively. }
    \label{fig1}
}
\noindent

We first note that any solution $\phi$ to the equations of motion
(\ref{eqmotn}) has its spectrum
in the interval $[0,s]$.
A proof of the equivalent result for scalar field theory on 
the Moyal plane is given by the 
authors of \cite{Durhuus:2001nj}. This proof carries over immediately 
to the fuzzy sphere so we do not reproduce it here.

If $\phi$ is any operator on $W_j$ we denote its matrix elements
$\br m|\phi |n\kt$ by $\phi_{m,n}$.
Before proceeding, we note that, in addition to rotational symmetry, 
the soliton equation (\ref{eqmotn}) is invariant under the transformation
\beq
\phi_{m,n} \rightarrow \phi_{-m,-n}
\eeq
which corresponds to reflection in the equator. Then, for example,
if $\phi$ is a diagonal soliton, so is $\psi$, where $\psi_m=\phi_{-m}$.

\section{Nonexistence at small $R$}\label{secnon}
In this section we show that, at fixed $j$, there are no solitons when the
radius $R$ is sufficiently small.  For convenience we define $\mu = R^2/2$. 
We define an inner product and a corresponding norm on hermitian matrices by
\beq
(\phi, \psi) = \Tr (\phi \psi ) \, , \;\;\; \norm{\phi}^2= \Tr \phi^2.
\eeq
Let $\phi$ be a solution to the soliton equation (\ref{eqmotn}) and let $I$
denote the unit matrix. 
We begin by showing that
\beq\label{bound}
\norm{\phi - rI}= \C O(\mu).
\eeq
The notation $f= \C O(\psi^n)$ will mean $\norm{f} =\C O (\norm{\psi}^n)$.

In order to prove (\ref{bound}) we note that any hermitian matrix 
$\phi$ can be split into two pieces
\beq
\phi = \phi_\parallel + \phi_\perp \label{split}
\eeq
with $\phi_\parallel$ proportional to the identity matrix, 
and $(\phi_\parallel , \phi_\perp)=0$ (so that $\Tr \phi_\perp=0$). 
The Laplacian operator (total angular momentum operator) 
in (\ref{eqmotn}) has eigenvalues $k(k+1)$ with 
$k=0,1, \cdots , 2j$. The unique eigenstate with $k=0$ is the identity 
matrix.  Substituting $\phi$ into (\ref{eqmotn}) gives
\beq
[J_i,[J_i,\phi_\perp]] + \mu V'(\phi ) =0
\eeq
so that
\beq\label{supineq}
\mu \norm{V'(\phi)} \geq 2 \norm{\phi_\perp}.
\eeq
Since any solution $\phi$ has its eigenvalues in the interval $[0,s]$, 
we find that
\beq
\norm{\phi_\perp} \leq c\mu ,
\eeq
where $c$ is a constant.
Next, we look at the trace condition
\beq
\Tr V'(\phi_\parallel+\phi_\perp)=0.
\eeq
Then
\beq
\Tr V'(\phi_\parallel) + \C O(\mu)=0
\eeq
and we conclude that for small values of $\mu$ 
\beq
\phi_\parallel=x_0 I +\C O(\mu),
\eeq
where $x_0$ is a zero of $V'$, and so  
\beq
\phi=x_0 I +\C O(\mu).
\eeq
Next, we diagonalise $\phi$ and discover eigenvalues 
$\phi_m=x_0 +\C O(\mu)$. Since every eigenvalue lies in the range 
$[0,s]$, the trace condition implies that there must be some eigenvalues 
smaller than $r$ and some larger than $r$. So by taking $\mu$ small 
enough, we must have $x_0=r$ and this proves 
equation (\ref{bound}).

If we now write
\beq
\phi=rI +\chi , \, \;\;\;\;\; \chi= \chi_\parallel + \chi_\perp ,
\eeq
using the splitting \rf{split}, 
and substitute into the soliton equation (\ref{eqmotn}), we obtain similarly
\beq\label{compare}
\mu \norm{V'(\phi)} \geq 2 \norm{\chi_\perp}.
\eeq
On the other hand we have
\bea
\norm{V'(\phi)}^2 &=& \Tr V'(rI+\chi)^2 \nonumber \\
&=& V''(r)^2 \norm{\chi}^2 + \C O (\chi^3) \\
&=&V''(r)^2 (\norm{\chi_\perp}^2 +\norm{\chi_\parallel}^2) + \C O (\chi^3). \nonumber
\eea
The trace condition gives 
\beq
\Tr V'(\phi)= \Tr (\chi_\perp + \chi_\parallel) V''(r) + \C O (\chi^2)
\eeq
and since $\Tr \chi_\perp=0$,
\beq
\norm{\chi_\parallel}^2 = \C O (\chi^4)
\eeq
so
\beq
\norm{V'(\phi)}^2=V''(r)^2 \norm{\chi_\perp}^2 + \C O (\chi^3).
\eeq
Comparing this with (\ref{compare}) we find
\beq
\mu^2 (V''(r)^2 \norm{\chi_\perp}^2 + \C O (\chi^3)) \geq 4\norm{\chi_\perp}^2.
\eeq
Noting that $\norm{\chi_\perp}^2= \norm{\chi}^2-\norm{\chi_\parallel}^2=
\norm{\chi}^2 (1+ \C O(\chi^2))$ we obtain finally
\beq
\mu^2 (V''(r)^2  + \C O (\chi^3)) \geq 4.
\eeq
If $\mu$ is small enough, this is impossible.

Recall that at fixed $j$ the radius $R$ determines the non-commutativity
parameter $\sigma$ through (\ref{ncparameter}).  We have thus established
the non-existence of scalar solitons at sufficiently weak non-commutativity
on the sphere.  This is certainly in line with previous results on 
the Moyal plane \cite{Durhuus:2000uz} but this was not guaranteed 
{\it a priori\/} given that Derrick's theorem only holds on the 
commutative plane but not on the sphere.

\section{Existence close to critical values of $R$}\label{seccrit}
In this section we establish the existence of solutions to the diagonal 
soliton equations (\ref{eq1}) and (\ref{eq2}) at certain critical values of
the radius $R$.  When we perform our stability analysis in 
section~\ref{sec:stab}, these solutions will all turn out to be unstable.
They do, however, provide insight into the structure of the field 
equations and we include them before moving on to the construction
in section~\ref{largetheta}, which yields stable solutions for sufficiently
large $R$.

We think of $\phi_{-j}$ as the first entry in the matrix
and then use the first order difference equation (\ref{eq1}) to find
each $\phi_m$ for $m>-j$ as a function of $\phi_{-j}\equiv x$. 
In order to discover whether a particular value of $x$ gives a soliton 
solution, we only need to check the constraint (\ref{eq2}).
This is equivalent to looking for zeros of the function 
\beq\label{eqg}
g_\mu(x)=\sum_{i=-j}^j V'[\phi_i (x)]
\eeq
where $\mu=R^2/2$.
We know that there are (trivial) solutions to the soliton equations at 
$x=0,r,s$ so $g_\mu(x)$ must have zeros at these points. We also know that 
when $\mu$ is small, $g_\mu(x)$ has no other zeros. Also, if $x<0$, 
the $\phi_m(x)$ get locked into the negative region of $V'$ and so 
$g_\mu(x)<0$. Similarly, when $x>s$, $g_\mu(x)>0$.  

Assuming that solitons do exist at some values of $\mu$, we can
imagine increasing $\mu$ to the critical point $\mu_c$ 
at which they come into existence. Imagine plotting a graph of 
$g_\mu(x)$ against $x$. As we move through the critical value of 
$\mu$, $g_\mu(x)$ will brush and then cut through the $x$ axis.   
This will happen at $\mu_c, x_c$ satisfying
\beq
g_{\mu_c}(x_c)= g_{\mu_c}'(x_c)=0,
\eeq
and by the continuity of $g_\mu(x)$, we will generically 
pick up a pair of new solitons.

In order to find the critical points, one in general needs to solve 
a nonlinear equation.  There is, however, a series of critical points 
with $x_c=r$ which we can determine analytically.

Setting $\phi_{-j}=x=r$, we solve the soliton equations and find
$\phi_m=r$ for all $m$, so that $g_\mu(r)=0$. Differentiating
Eq.\ (\ref{eqg}) with respect to $x$ and putting $x=r$ leads to the constraint
\beq\label{critconstraint}
g_\mu'(r)= V''(r) \sum_{i=-j}^j \phi_i'(r)=0.
\eeq
We can differentiate (\ref{diffeq}) to obtain a difference equation for 
the $\phi_i'(r)$, with initial value $\phi'_{-j}(r)=1$
\beq\label{critdiffeq}
\alpha_m (\phi'_{m+1}(r)-\phi'_{m}(r)) -\alpha_{m-1} (\phi_{m}'(r)-
\phi_{m-1}'(r))= \mu V''(r) \phi'_m(r).
\eeq
Let us define a diagonal matrix $u$ with $u_{ii}= \phi'_i(r)$ 
(note that, although we have not explicitly written it, the matrix $u$ 
is a function of $\mu$, inherited from the definition of the $\phi_i$). 
As long as $\mu \neq 0$, we can combine (\ref{critconstraint}) 
with (\ref{critdiffeq}) to obtain an equivalent matrix equation
\beq
[J_i,[J_i,u]] + \mu V''(r ) u =0.
\eeq
The Laplacian has eigenvalues $k(k+1)$, with $k=0,1,2, \cdots , 2j$, and
there is precisely one eigenstate represented by a diagonal matrix
$e_k$ for each 
eigenvalue. Therefore the critical values of $\mu$ at $x_c=r$ are
\beq\label{critpoints}
\mu_c = {-k(k+1) \over V''(r)} \, , \;\;\; k=1,2, \cdots , 2j,
\eeq
and at the critical point we have 
\beq\label{uc}
u(\mu_c) =ae_k, ~a\neq 0.
\eeq
We note that the number of critical points increases with $j$ but the
lowest critical value of $\mu$ (and therefore of $R$) is independent
of $j$.

The picture that we have is the following. As we increase $\mu$
towards a critical point, the function $g(x)$ flattens at $x=r$, and as
we move through the critical point, we expect $g$ to start to cut through the
$x$-axis on either side of $x=r$ so that we gain two new solitons. 
Alternatively, it is possible that this process will precisely correspond 
with another pair of solitons moving in towards the critical point leading 
to a cancellation so that the net effect is the loss of a pair of solitons. 
In any case, when $g'(r)$ changes sign, we must either gain 
or lose two soliton solutions since $g(x)$ is continuous.

There is an additional possibility which we cannot yet discount. Even though $g'(r)=0$ at the critical value of $\mu$, it is possible that it may not change sign as we move through the critical value. This can only be if
\beq\label{contradict}
{\partial \over \partial \mu} \, g'_\mu(r)|_{\mu=\mu_c} = 0.
\eeq
If we assume (\ref{contradict}) is true, then by a very similar argument to the previous, we obtain the matrix equation
\beq
[J_i,[J_i,\dot{u}(\mu_c)]] + V''(r ) u(\mu_c) + \mu_c V''(r) \dot{u}(\mu_c) =0
\eeq
where $\dot{u}$ denotes $u$ differentiated with respect to $\mu$. As before, we can expand $u(\mu)$ in terms of the eigenstates of the Laplacian $u(\mu)= \sum_{l=0}^{2j} u_l(\mu) e_l$. We are studying a critical point satisfying $\mu_c V''(r) =-k(k+1)$ for some fixed $k$, so let us pick out the coefficient of $e_k$:
\beq
0=k(k+1)\dot{u}_k(\mu_c) + V''(r ) u_k(\mu_c) + \mu_c V''(r) \dot{u}_k(\mu_c) = V''(r ) u_k(\mu_c).
\eeq
But by (\ref{uc}), $u(\mu_c)=u_k(\mu_c)e_k \neq 0$, so this is a contradiction.

The conclusion is that whenever $\mu$ moves through one of the critical points (\ref{critpoints}), we either gain or lose two solitons. As a corollary, there is a range of $\mu$ close to each of the critical points for which solitons exist.

To illustrate, figure \ref{fig2} shows an example plot of the zeros of $g_\mu(x)$ for $j={3 \over 2}$ and a potential with $V'(u)=u(u-1.2)(u-2.0)$. We see that there are many critical points in addition to those we have calculated. In most cases, soliton solutions are created as we increase $\mu$, but they are also occasionally lost. In some cases, the critical points occur as bifurcations of existing solitons, but in most cases a pair of solitons appears (or disappears) where there was none before. Nevertheless,  we call this the bifurcation diagram.
\FIGURE[hbt]{
    \psfrag{mu}{$\mu$}
   \psfrag{x}{$x$} 
    \includegraphics[width=10cm]{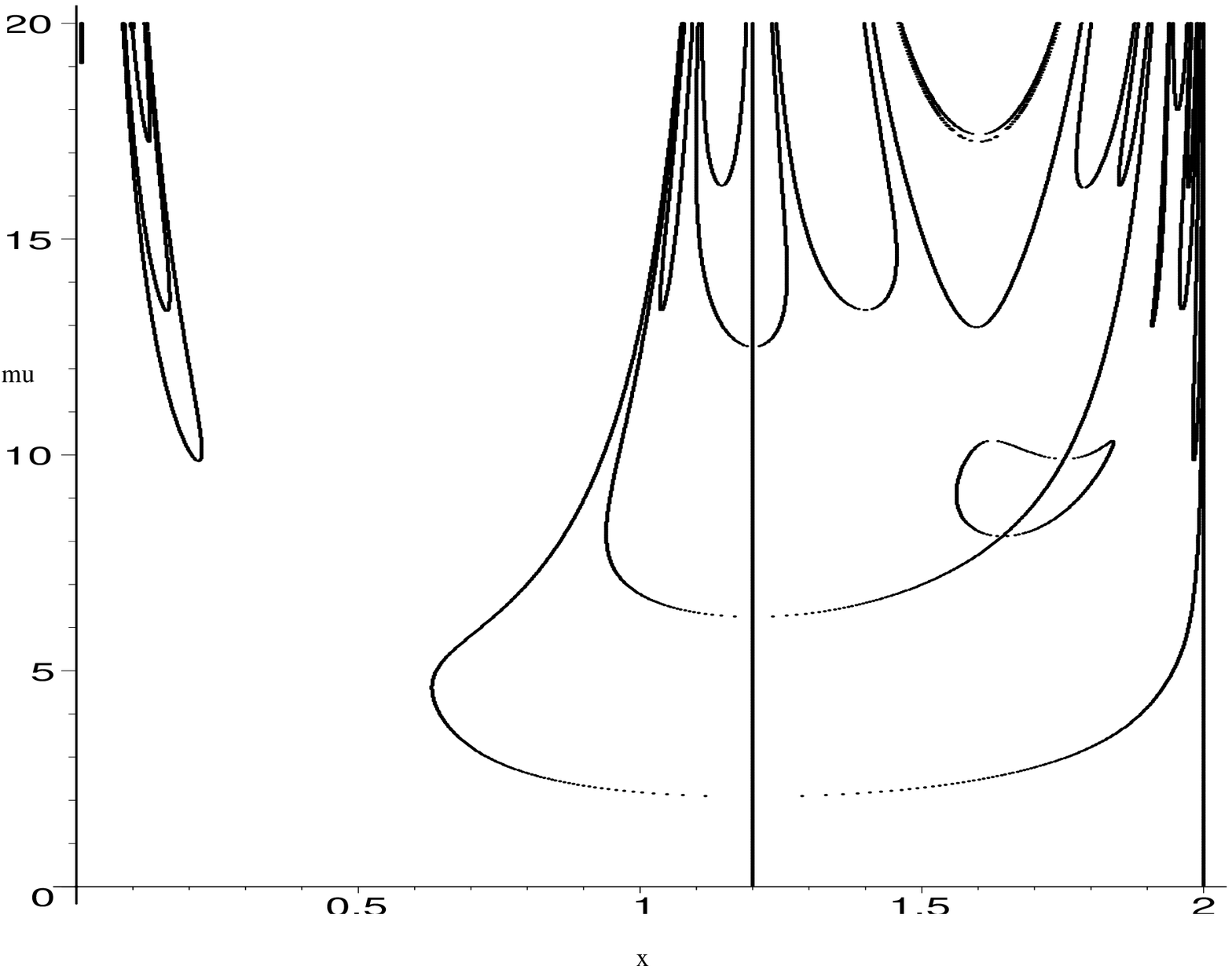}
    \caption{Diagram showing the zeros of $g_\mu(x)$ for $j={3 \over 2}$.}
    \label{fig2}
}
We have observed that if one increases $j$, the complexity of the bifurcation diagram increases, and the number of solitons at sufficiently large $\mu$ increases. 

\section{Diagonal solitons}
\label{largetheta}
We now show, for any given value of $j$, that the diagonal soliton equations 
(\ref{eq1}) and (\ref{eq2}) have a solution provided $R$ is large enough.  
In section~\ref{sec:stab} we will further show that this construction leads to
stable solutions.  As $j$ increases, we will need to increase $ R^2$ linearly, 
so let us write $R^2/2= \theta j$ at the outset. When we come to consider the 
planar limit, it will also be convenient to rewrite the matrix elements of 
$\phi$
\beq
\phi=\sum_{m=-j}^{j} \phi_m |m \rangle \langle m| = \sum_{q=0}^{2j} \psi^{(j)}_q |-j+q \rangle \langle -j+q|
\eeq
so that $\psi^{(j)}_q=\phi_{-j+q}$. Then the equations of motion (\ref{eq1}), (\ref{eq2}) become
\beq
\label{shiftelts1}
\psi^{(j)}_{q+1}-\psi^{(j)}_q = {\theta \over 2(q+1)(1-q/2j)} \sum_{k=0}^q V^\prime (\psi^{(j)}_k)\, , \;\;\; q=0,\cdots , 2j-1,
\eeq
\beq
\label{shiftelts2}
\sum_{k=0}^{2j} V^\prime(\psi^{(j)}_k) =0.
\eeq
We have included an index $j$ in $\psi^{(j)}_q$ to remind ourselves that we are working with the fuzzy sphere, $M_j$, which is the $j$-th representation of $su(2)$.

We will see that solutions do exist for sufficiently large $\theta$ by adapting the proof given in \cite{Durhuus:2000uz} for the Moyal plane.\\

\noindent
\begin{bf}{Theorem 1}\end{bf}
\newline
{\it There is a constant, $\theta_c$, independent of $j$ such that when $\theta > \theta_c$ equations (\ref{shiftelts1}), (\ref{shiftelts2}) have a non-trivial solution.}
\\

\noindent
To see this, suppose $V^\prime$ has its first local maximum in the domain $[0,s]$ at $u$ and its final local minimum at $v$ (see figure \ref{fig1}). We begin by choosing 
\beq
\theta > {-2v \over V'(v)}.
\eeq
Then setting $q=0$ in (\ref{shiftelts1}), we have 
\beq
\psi^{(j)}_1=\psi^{(j)}_0 + {\theta \over 2} V'(\psi^{(j)}_0),
\eeq
so there is a unique $x$ with $v<x<s$ such that if $\psi^{(j)}_0=x$ then $\psi^{(j)}_1=0$, and this $x$ is independent of $j$. As $\theta$ increases, $x$ moves towards $s$ independently of $j$.
Then we can choose $\theta$ sufficiently large that 
\beq\label{con}
\absval{V'(x)}< V'(u).
\eeq
This is the final constraint on $\theta$, so our critical value $\theta_c$ will indeed be independent of $j$.

Now if we increase $\psi^{(j)}_0$ from $x$ towards $s$, $\psi^{(j)}_1$ increases to $s$. Therefore, we can choose $y$ such that $x<y<s$ and when $\psi^{(j)}_0=y$ then $\psi^{(j)}_1=u$.

Using (\ref{shiftelts1}), each $\psi^{(j)}_0$ defines a sequence $\psi^{(j)}_k$. We next define the set
\beq
\label{setA}
A^{(j)} = \left\{ \psi^{(j)}_0 \in [x,y) : \exists q, 0\leq q \leq 2j, \sum_{k=0}^q V'(\psi^{(j)}_k)>0 \right\}.
\eeq
We see from (\ref{shiftelts1}) that if $\psi^{(j)}_0$ leads to a sequence $\psi^{(j)}_k$ which is not monotonically decreasing, then $\psi^{(j)}_0$ is in $A^{(j)}$. In addition there may be a monotonically decreasing sequence for which $\psi^{(j)}_0$ is in $A^{(j)}$, if $\sum_{k=0}^{2j} V'(\psi^{(j)}_k)>0$.

We note that $x \notin A^{(j)}$, since the sequence which starts at $x$ jumps first to $0$, and then gets locked into the negative region so continues to decrease. On the other hand, using (\ref{con}), we see that $y-\epsilon \in A^{(j)}$ as long as $\epsilon$ is sufficiently small. So $A^{(j)}$ is not empty. Since $V'$ is continuous, we also note that $A^{(j)}$ is open.

Following \cite{Durhuus:2000uz}, we choose
\beq
\psi_0^{(j)}= \inf \, A^{(j)}.
\eeq
Then $\psi_0^{(j)} \notin  A^{(j)}$ since $A^{(j)}$ is open, and so
\beq
\label{cont}
\sum_{k=0}^q V'(\psi^{(j)}_k) \leq 0 \, , \;\;\; q=0, \cdots ,2j.
\eeq
But since $\psi_0^{(j)}$ is the infimum of $A^{(j)}$, there must by continuity be at least one $q$ such that
\beq
\sum_{k=0}^q V'(\psi^{(j)}_k) = 0
\eeq
and we choose $\widetilde{q}$ to be the smallest such $q$.

Let us suppose $\widetilde{q}<2j$. Then since $\sum_{k=0}^{\widetilde{q}-1} V'(\psi^{(j)}_k) < 0$, we have certainly $V'(\psi^{(j)}_{\widetilde{q}})>0$. Using (\ref{shiftelts1}), $\psi^{(j)}_{\widetilde{q}+1}=\psi^{(j)}_{\widetilde{q}}$ and therefore $\sum_{k=0}^{\widetilde{q}+1} V'(\psi^{(j)}_k) > 0$, which contradicts (\ref{cont}).
We deduce that $\sum_{k=0}^{2j} V'(\psi^{(j)}_k) = 0$, and so we have completed the construction of a solution to the equations (\ref{shiftelts1}), (\ref{shiftelts2}) for sufficiently large $\theta$. This solution has the property that the sequence $\psi^{(j)}_k$ is monotonically decreasing.\\

We can extend the proof given above (in the same manner as in \cite{Durhuus:2001nj}) to construct further soliton solutions. This time, we fix an integer $N$ with $0<N \leq 2j$ and require
\beq
\theta > {-2v N \over V'(v)}.
\eeq
Then there is a unique $x^{(j)}$ such that if $\psi^{(j)}_0=x^{(j)}$, then
\beq
\psi^{(j)}_0 > \psi^{(j)}_1 > \cdots > \psi^{(j)}_{N-1} \geq v
\eeq
and $\psi^{(j)}_{N}=0$. 

Only in the case of $N=1$ which we covered before is $x^{(j)}$ independent of $j$. However, we note that $x^{(j)} \geq x^{(j+1)}$, and define $\lim_{j \rightarrow \infty} x^{(j)} = x^{(\infty)} \in (v,s)$. If we take the limit of equation (\ref{shiftelts1}) as $j \rightarrow \infty$, we see that as we increase $\theta$ to infinity, $x^{(\infty)}$ increases to $s$. This means that we can make each $x^{(j)}$ arbitrarily close to $s$ by increasing $\theta$ \begin{it}independently of\end{it} $j$.

We next take $\theta$ sufficiently large that 
\beq
N \absval{V'(x^{(\infty)})} < V'(u). 
\eeq
This is the final constraint on $\theta$ and we note that it depends on $N$, but not $j$. We choose $y^{(j)}> x^{(j)}$ such that if $\psi^{(j)}_0=y^{(j)}$ then $\psi^{(j)}_{N+1}=u$. Replacing $x$ and $y$ with $x^{(j)}$ and $y^{(j)}$ in the definition (\ref{setA}) of $A^{(j)}$, the proof now proceeds as before.

We have constructed for each $N=1,2,\cdots, 2j$ a soliton with the property that the $\psi^{(j)}_q$ form a monotonically decreasing sequence. The first $N$ eigenvalues lie in the interval $(v,s)$ and the others lie in $(0,u)$. In particular, we note that each $\psi^{(j)}_q$ satisfies $V''(\psi^{(j)}_q)>0$. In the large $\theta$ limit, the first $N$ eigenvalues tend to $s$ and the others to zero.

\section{Stability}\label{sec:stab}
In this section, we investigate the stability of diagonal solitons. We will 
show that a necessary condition for stability is that the eigenvalues form a 
monotonic sequence. Next, we specialise to values of $\mu$ greater than some 
critical value $\mu_1$ depending on the potential $V$ and on $j$. Then another 
necessary condition for stability is that the eigenvalues lie in the regions 
close to the troughs of $V$ in which $V''>0$. Together, these two conditions 
are also sufficient for stability, so that the solitons we constructed in 
section \ref{largetheta} are stable for sufficiently large $\mu$. Finally, we 
show that the only deformations of a stable soliton which leave its energy 
invariant are rotations.

The tools for analysing stability have been set up and discussed in detail in \cite{Durhuus:2001nj} in the context of the Moyal plane (and see also \cite{Jackson:2001iy} in which stability under radial perturbations was considered). The stability functional is defined
\beq
\Sigma (\omega)= \left. \half {d^2 \over d \epsilon ^2} S(\phi + \epsilon \omega ) \right|_{\epsilon=0}
\eeq
where $\phi$ is a soliton solution. If $\Sigma$ is nonnegative then the soliton $\phi$ is said to be stable. We will find that $\Sigma$ always has zero eigenvalues because of the rotation symmetry of the action.  

Substituting for the action, we can rewrite the stability functional as
\beq\label{stabfnl}
\Sigma(\omega)= J(\omega) + \left. \mu {d^2 \over d \epsilon ^2} \Tr V(\phi + \epsilon \omega ) \right|_{\epsilon=0}
\eeq
where $J(\omega)= \Tr[J_i,\omega][\omega,J_i]$ and $ \mu= { R^2 /2}$. Using standard non-degenerate perturbation theory (see \cite{Durhuus:2001nj}), this can be written
\beq\label{stab}
\Sigma(\omega)= J(\omega) + 2\mu \sum_{m<n} \absval{\langle n| \omega |m \rangle}^2 {V'(\phi_m)-V'(\phi_n) \over \phi_m - \phi_n} + \mu \sum_{n=-j}^j \absval{\langle n| \omega |n \rangle}^2 V''(\phi_n)
\eeq
as long as $\phi_m \neq \phi_n$ for all $m$, $n$. If there are degenerate eigenvalues $\phi_m=\phi_n$, we can use the fact that the expression (\ref{stabfnl}) is continuous in $\phi$, and obtain the correct formula by taking the limit of (\ref{stab}) as the eigenvalues approach each other. In this way, we see that we must make the substitution 
\beq
{V'(\phi_m)-V'(\phi_n) \over \phi_m - \phi_n} \rightarrow V''(\phi_m)
\eeq
in equation (\ref{stab}).

Examining the equation of motion (\ref{eqmotn}), we see that the eigenvalues must lie within $\C O(\mu^{-1})$ of the zeros of $V'$. If any $\phi_m$ is close to $r$, then since $V''(r)<0$, the last term in (\ref{stab}) can be made large and negative, so the soliton is unstable. So choosing $\mu$ sufficiently large, our first necessary condition for stability is that each $\phi_m$ lies in the region $ [0,u]$ or $[v,s]$ (see figure \ref{fig1}). This condition simply says that the eigenvalues must all lie in the regions close to the troughs of $V$ in which $V''>0$.

We can write
\beq
J(\omega)= \sum_{m,n} \absval{ \sqrt{\alpha_{n-1}} \langle n| \omega |m \rangle - \sqrt{\alpha_{m-1}} \langle n-1| \omega |m-1 \rangle}^2 +(n-m)^2 \absval{\langle n| \omega |m \rangle}^2
\eeq
and note that $\Sigma$ is a quadratic form in the matrix elements of $\omega$, and these matrix elements are only coupled along diagonals, which we label by an integer $k$. We dealt with the case $k=0$ in the previous paragraph, and so the stability problem can be reduced to studying the reduced functional
\bea
\Sigma_k (\omega) &=& \sum_{\absval{n-m}=k} \absval{ \sqrt{\alpha_{n-1}} \langle n| \omega |m \rangle - \sqrt{\alpha_{m-1}} \langle n-1| \omega |m-1 \rangle}^2 \nonumber \\
&&+ \sum_{\absval{n-m}=k} (n-m)^2 \absval{\langle n| \omega |m \rangle}^2 \nonumber \\
&&+2\mu \sum_{n-m=k} \absval{\langle n| \omega |m \rangle}^2{ V'(\phi_m)-V'(\phi_n) \over \phi_m - \phi_n}
\eea
for each $k>0$.

We fix $k$ and define $x_n=\langle n+k| \omega |n \rangle$ for $n=-j, \cdots, j-k$, so that $x_n^*=\langle n| \omega |n+k \rangle$. Then we can write
\beq
\Sigma_k (\omega)= 2 \sum_{m,n} q_{mn} x_m x_n^*
\eeq
where the only non-zero elements of the quadratic form $q$ are
\bea
q_{nn}&=&\alpha_{n}+\alpha_{n+k-1}+k(k-1) +\gamma_n \label{qdiag}\\
q_{n,n-1}&=&-\sqrt{\alpha_{n-1}\alpha_{n+k-1}}
\eea
and where
\beq
\gamma_n= \mu {V'(\phi_{n+k})-V'(\phi_{n}) \over \phi_{n+k}-\phi_{n}} \, , \;\;\; \phi_n \neq \phi_{n+k}
\eeq
\beq
\gamma_n= \mu V''(\phi_n)  \, , \;\;\; \phi_n = \phi_{n+k}.
\eeq
To obtain (\ref{qdiag}), we used the identity
\beq
\half (\alpha_{n+k-1}+\alpha_n +\alpha_{n-1}+\alpha_{n+k})= \alpha_{n}+\alpha_{n+k-1}-k.
\eeq
We will follow \cite{Durhuus:2001nj} to use elementary row and column operations to find a new diagonal quadratic form, $C$, which has the same numbers of positive, negative and zero eigenvalues as $q$. Specifically, we define the diagonal elements of $C$ inductively by
\bea
C_{-j}&=&q_{-j,-j}\\
C_n&=&q_{n,n}- {q_{n,n-1}^2 \over C_{n-1}} \, , \;\;\; (n=-j+1,\cdots ,j-k).
\eea
We are now ready to prove another necessary condition for stability.\\

\noindent
\begin{bf}{Theorem 2}\end{bf}
{\it A necessary condition for stability is that the eigenvalues $\phi_n$ form a monotonic sequence.}\\

\noindent
We note first that we do not need to assume $\mu$ is large.\\
We examine $\Sigma_k(\omega)$ with $k=1$. We can use (\ref{diffeq}) to eliminate $V'(\phi_n)$ and $V'(\phi_{n+1})$ from $\gamma_n$, and find
\beq
q_{nn}= \alpha_{n+1} {\phi_{n+2}-\phi_{n+1} \over \phi_{n+1}-\phi_{n}} + \alpha_{n-1} {\phi_{n}-\phi_{n-1} \over \phi_{n+1}-\phi_{n}},
\eeq
as long as $\phi_n \neq \phi_{n+1}$.
Let us assume first that there are no such pairs of consecutive degenerate eigenvalues. Then it is very easy to check by induction that
\beq\label{Ck1}
C_n= \alpha_{n+1} {\phi_{n+2}-\phi_{n+1} \over \phi_{n+1}-\phi_{n}}
\eeq
for all $n=-j, \cdots, j-1$. If the $\phi_n$ are not monotonic, at least one of the $C_n$ must be negative.

For the degenerate case, assume that the first pair of consecutive degenerate eigenvalues is $\phi_p=\phi_{p+1}$. Then we obtain
\beq
C_n= \alpha_{n+1} {\phi_{n+2}-\phi_{n+1} \over \phi_{n+1}-\phi_{n}} \, , \;\;\; n \leq p-1,
\eeq
and in particular, $C_{p-1}=0$ so that this inductive reduction fails at $n=p-1$. However, we can use the $(p-1)$-th row and column to further eliminate and bring the matrix into block diagonal form, and then $C_p$ and $C_{p+1}$ are given by the eigenvalues of
\beq
\left(
\begin{array}{cc}
0&-\sqrt{\alpha_{p-1}\alpha_{p-2}}\\
-\sqrt{\alpha_{p-1}\alpha_{p-2}}&2\alpha_p + \mu V''(\phi_p)
\end{array}
\right)
\eeq
one of which is always negative. This completes the proof of Theorem 2.\\

In equation (\ref{Ck1}), we note that $C_{j-1}$ is zero (since $\alpha_{j}=0$). This zero eigenvalue in the stability functional comes about because we can rotate the soliton, and obtain a new solution with the same energy. In general, if there is a family of soliton solutions $\phi(\nu)$ depending smoothly on $\nu$, with $\phi(0)=\phi$, then necessarily there exists an $\omega \neq 0$ with $\Sigma (\omega)=0$. To see this, take $\omega=d\phi / d \nu |_{\nu=0}$. We can always choose a parameterization such that $\omega \neq 0$. Then, expanding $\phi$ in terms of some basis for the matrices, $\phi= \phi^i T^i$, we have
\beq
0=\half \left. {d^2 \over d \nu^2}S[\phi(\nu)]\right\vert_{\nu=0}= \half \left.{\pa^2S \over \pa \phi^i \pa \phi^j} \right\vert_{\nu=0} \omega^i \omega^j + {\pa S \over \pa \phi^i} \left.{d^2 \phi^i \over d \nu^2}\right|_{\nu=0}
\eeq
The first term is the stability functional, whilst the second term is zero by the equation of motion, so we obtain $\Sigma(\omega)=0$.

Since the rotations are a symmetry of the action, they will lead to zero eigenvalues of the stability functional. The rotations are a three parameter group, but one of these (rotation about the $z$-axis) leaves a diagonal soliton invariant. This means we would expect to find two zero eigenvalues in the stability functional corresponding to the rotations. The quadratic form $q$ acts on a complex space, and so the zero eigenvalue $C_{j-1}$ corresponds to two zero eigenvalues on a real space. Our task now is to show that this zero eigenvalue corresponds precisely to the rotations. 

We do this by showing that any hermitian matrix $\C X \in M_j$ can always be rotated into a canonical form.\footnote{We recall this notation from section \ref{secfuzsphere}. The fuzzy sphere is the algebra $M_j$ of $(2j+1) \times (2j+1)$ complex matrices.} Then we can restrict our space to matrices of this form, and check that the zero eigenvalues disappear. Specifically, given any hermitian matrix $\C X_{pq}$, $(-j \leq p,q \leq j)$, we can always use a rotation to set $\C X_{j,j-1}= \C X_{j-1,j}=0$, and we shall restrict to this rotation fixed space of matrices.

To see that this is possible, it is helpful to use the coherent state representation for the algebra. The formulae that we use can be found in \cite{Peremelov}. A general rotation generated by $J_1$ and $J_2$ can be re-written in \begin{it}normal form\end{it}
\beq
\exp(i \theta_1 J_1 + i \theta_2 J_2) = \exp(\zeta J_+) \exp(\eta J_3) \exp(-\overline{\zeta}J_-) \equiv D(\zeta)
\eeq
where, defining $\theta=\theta_1+i\theta_2$, we have $\zeta=i\absval{\theta}^{-1}\tan (\absval{\theta}/2)\, \theta$ and $\eta=\log(1+\absval{\zeta}^2)$. It can also be written in {\it antinormal form}
\beq
D(\zeta)= \exp(-\overline{\zeta}J_-)\exp(-\eta J_3)\exp(\zeta J_+).
\eeq
The coherent state is defined
\beq
|z\rangle = D(z) |j,-j\rangle,
\eeq
and is automatically normalised $\langle z | z\rangle=1$. 

The operators $D$ satisfy
\beq\label{cohphase}
D(\zeta_1) D(\zeta_2)=D(\zeta_3) \exp(i \Phi J_3)
\eeq
where $\zeta_3$ and $\Phi$ are respectively complex and real numbers depending on $\zeta_1$ and $\zeta_2$. In particular, $D(\zeta) D(-\zeta)=1$ as can be checked by multiplying the normal and antinormal forms.

Given a hermitian matrix $\C X \in M_j$, we define the covariant symbol \cite{Berezin}
\beq
\C X(z,\zbar)= \langle z | \C X | z \rangle,
\eeq
which, in terms of matrix elements in the standard basis, is
\beq\label{covtelts}
\C X(z,\zbar) = {1 \over (1+ \absval{z}^2)^{2j}} \sum_{m,n=-j}^{j} \C X_{mn} \, {\zbar^{m+j} z^{n+j}} \sqrt{{2j \choose m+j} {2j \choose n+j}}.
\eeq
The covariant symbol is one way of mapping the fuzzy sphere algebra $M_j$ to an algebra of functions on the sphere (with noncommutative $*$-product). Here the sphere is given in stereographic coordinates, and we note that if the matrix $\C X$ is diagonal, $\C X(z,\zbar)$ depends only on $\absval{z}$, so that it is indeed rotationally invariant about the north-south axis.

The function $\C X(z,\zbar)$ must have at least one stationary point, $z_0$, in $\absval{z}< \infty$. One can quickly check using (\ref{cohphase}) that applying a rotation $\C X \rightarrow \widetilde{\C X}= D^\dagger(\zeta) \C X D(\zeta)$ leads to a new covariant symbol $\widetilde{\C X}(z,\zbar)= \C X(z',\zbar')$ for some $z'$ depending on $z$ and $\zeta$. In particular, we can apply the rotation $D(-z_0)$ to get a covariant symbol with a stationary point at $z=0$. Then differentiating (\ref{covtelts}) at $z=0$ gives 
\beq
\widetilde{\C X}_{j,j-1}=\widetilde{\C X}_{j-1,j}=0
\eeq
which is our desired rotation fixing condition.

We can now deal with the zero eigenvalue $C_{j-1}$. We restrict to the rotation fixed space of matrices $\C X$, with $\C X_{j,j-1}= \C X_{j-1,j}=0$. In this case, the above analysis works identically except that $C_{j-1}$ no longer appears. This shows that the zero eigenvalue corresponds precisely to the rotations. In particular, when the eigenvalues form a monotonic sequence, $\Sigma_1$ is positive definite on the rotation fixed space. 

We now go on and consider $\Sigma_k$ for $k>1$.\\

\noindent
\begin{bf}{Theorem 3}\end{bf}
{\it Suppose $\phi$ is a diagonal soliton, and the eigenvalues $\phi_m$ form a strictly monotonic sequence, and suppose $\mu$ is sufficiently large. Assume further that every eigenvalue $\phi_m$ has $V''(\phi_m)>0$. Then the quadratic forms $\Sigma_k$ are positive definite on the rotation fixed space.}\\

\noindent
We have already covered the cases $k=0$ and $k=1$, so we restrict now to $k>1$.
Since $\mu$ is large, the condition $V''(\phi_m)>0$ means that the eigenvalues can lie close to $0$ and $s$, but not $r$.
We choose, without loss of generality, the sequence of eigenvalues to be decreasing (rather than increasing). Then there exists $N$ such that 
\bea
\phi_{-j}, \cdots , \phi_{-j+N-1} &=&s - \C O (\mu^{-1})\\
\phi_{-j+N}, \cdots , \phi_{j} &=& \C O (\mu^{-1})
\eea
that is the first $N$ eigenvalues are close to $s$, and the remaining are close to $0$. We can use (\ref{eq1}) to compute the $\phi_m$ to first order in $\mu^{-1}$. We see immediately that if $\phi_m=s-\C O(\mu^{-p})$, then $\phi_{m+1}=s-\C O(\mu^{-p+1})$ for $m \leq -j+N-1$. Then we can write
\bea
-s+ \C O(\mu^{-1}) &=& \phi_{-j+N} - \phi_{-j+N-1}={\mu \over \alpha_{-j+N-1}}\sum_{k=-j}^{-j+N-1} V'(\phi_k) \nonumber \\&=& {\mu \over \alpha_{-j+N-1}}V'(\phi_{-j+N-1}) + \C O(\mu^{-1})\nonumber \\&=&  {\mu \over \alpha_{-j+N-1}} V''(s) (\phi_{-j+N-1}-s)+ \C O(\mu^{-1})
\eea 
giving
\beq
\phi_{-j+N-1}=s- {s \, \alpha_{-j+N-1} \over V''(s)} \mu^{-1} + \C O(\mu^{-2})
\eeq
and
\beq
\phi_{-j+N-q}=s-\C O(\mu^{-q}) \, , \;\;\; q \geq 2.
\eeq
Similarly,
\beq
\phi_{-j+N}={s \, \alpha_{-j+N-1} \over V''(0)} \, \mu^{-1} + \C O(\mu^{-2})
\eeq
and
\beq
\phi_{-j+N-1+q}=\C O(\mu^{-q}) \, , \;\;\; q \geq 2.
\eeq
We can use these results to evaluate the $\gamma_m$ to first order, and find bounds on the $C_m$.

First, when $m+k \leq -j+N-1$, we have $\gamma_m=\mu V''(s) + \C O(1)$ giving
\beq
C_m=\mu V''(s) + \C O(1) \, , \;\;\; m \leq -j+N-k-1.
\eeq
Next, if $m=-j+N-k$, we find $\gamma_{-j+N-k}= -\alpha_{-j+N-1}+\C O(\mu^{-1})$, giving
\beq
C_{-j+N-k}=\alpha_{-j+N-k}+k(k-1) +\C O(\mu^{-1}).
\eeq 
Then for $-j+N-k< m< -j+N-1$, we have $\gamma_m = \C O(\mu^{-1})$, and we can check inductively that
\beq
C_m > \alpha_m +k(k-1) +\C O(\mu^{-1}) \, , \;\;\; -j+N-k< m< -j+N-1.
\eeq
When $m=-j+N-1$, we have $\gamma_{-j+N-1}= -\alpha_{-j+N-1}+\C O(\mu^{-1})$ giving
\beq
C_{-j+N-1}> k(k-1) +\C O(\mu^{-1}).
\eeq
Finally, if $m>-j+N-1$, we have $\gamma_m = \mu V''(0) + \C O(1)$, and so
\beq
C_m = \mu V''(0) + \C O(1) \, , \;\;\; m>-j+N-1.
\eeq
This completes the proof of theorem 3, since every $C_n$ is positive at sufficiently large $\mu$.\\

We have shown that a soliton which satisfies the assumptions of theorem 3 has positive definite stability functional on the rotation fixed space, and so is stable. In particular, the solitons which we constructed in theorem 1 are stable when $\mu$ is sufficiently large. These solitons cannot be deformed without increasing their energy, because this would require the stability functional to have zero eigenvalues after fixing the rotations.

\section{Unstable multi-lump solutions}\label{secunstable}
In this section we consider solutions of the field equations that describe
separated lumps on the fuzzy sphere.  As mentioned in the Introduction, a
precise mathematical definition of a multi-soliton is lacking at finite 
non-commutativity but one can nevertheless look for solutions where the 
scalar field has nonvanishing values in well separated regions on the 
sphere.  Perturbative calculations at large non-commutativity indicate 
that such lumps attract each other \cite{Spradlin:2001ku,Vaidya:2001rf} and
therefore one would not expect to have solutions with multiple lumps at
arbritrary locations.  On symmetry grounds, however, one might anticipate 
that there could exist solutions with lumps in an unstable equilibrium. 
The simplest such solution would have two identical lumps, one at each
pole of the sphere.  Such a soliton would be rotationally invariant about
the $z$-axis but also invariant under reflection about the equatorial plane.  
It would thus be a diagonal solution with the eigenvalues satisfying
\beq
\phi_m= \phi_{-m}.
\eeq
If $j$ is a half integer, the equations (\ref{eq1}), (\ref{eq2}) reduce consistently under this assumption to
\beq
\phi_{m+1}-\phi_m={ R^2 \over 2  \alpha_m} \sum_{i=-j}^m V^\prime (\phi_i) \, , \;\;\; m=-j,\cdots , -{3 \over 2}
\eeq
with the constraint
\beq
\sum_{i=-j}^{-{1 \over 2}} V^\prime (\phi_i) =0.
\eeq
To see this, we use the property $\alpha_{-m}=\alpha_{m-1}$ to check that (\ref{eq1}) also holds for $m \geq \half$. When $j$ is an integer, the equations become
\beq
\phi_{m+1}-\phi_m={ R^2 \over 2 \alpha_m} \sum_{i=-j}^m V^\prime (\phi_i) \, , \;\;\; m=-j,\cdots , -1
\eeq
with the constraint
\beq
2\sum_{i=-j}^{-1} V^\prime (\phi_i) +V^\prime (\phi_0)=0.
\eeq
We can suitably adjust the set $A^{(j)}$ (equation (\ref{setA})) to the altered difference equation and constraint. Then applying the argument of 
section~\ref{largetheta} shows immediately that these soliton solutions do 
exist for sufficiently large $\theta$. We see from the stability analysis in 
section~\ref{sec:stab} that such solitons are unstable since the 
eigenvalues do not form a monotonic sequence.

We can construct a bifurcation diagram for these solutions which must be a subset of the bifurcation diagram for the full problem. Then we see that they correspond precisely with certain trajectories in the full bifurcation diagram. In particular, the diagonal eigenstates of $[J_i,[J_i, \; \cdot \; ] \, ]$ corresponding to eigenvalue $k(k+1)$ with $k$ even have the property $\phi_m=\phi_{-m}$, so every second bifurcation from $x=r$ corresponds to an unstable solution of this type.

By symmetry, one would expect to be able to go further, and exhibit unstable solitons with $n$ lumps evenly spaced around a great circle. At fixed $j$, this $n$ can be at most $2j$, as the fuzziness prevents resolution of more lumps. Here, we briefly report results for a specific example. We examine $j={3 \over 2}$ and seek a $3$-lump solution. We begin with a general diagonal matrix $P$, and a rotation through $2 \pi/3$ about the $x$-axis, $\C R=\exp(2 \pi iJ_x /3)$, and define a new matrix $Q$
\beq
Q= P +\C R P \C R^{-1} + \C R^{-1}PR
\eeq
which is invariant under rotation through $2\pi /3$ about the $x$-axis. In this case, the situation is simplified because all such matrices $Q$, and kinetic terms $[J_i,[J_i,Q]]$ are simultaneously diagonalisable, and two of the eigenvalues are always identical. One can then solve the equations of motion for the eigenvalues, and find several solutions when $\mu$ is chosen large enough. 

In the following section we take a limit of large $R$ and $j$ to blow up a small region close to the south pole into the Moyal plane. The multi-lump solutions constructed in this section will not survive this limit since any lump not sitting at the south pole will be moved to infinity. However, by symmetry one would expect the existence of an unstable solution consisting of an infinite chain of lumps finitely separated along a straight line in the Moyal plane, and its generalisation to a two-dimensional array of lumps.\footnote{In [16] such periodic solutions are constructed in the limit 
of large $\theta$.} Such solitons would of course have infinite energy.

\section{The planar limit}
\label{planar}
An additional motivation for studying the fuzzy sphere is that, by blowing up a small region close to, say, the south pole, there is a weak convergence of the algebra to that of the Moyal plane \cite{Chu:2001xi,Madore:1992bw}. This means that the fuzzy sphere may be used to cut off the infinite dimensional algebra and provide an infra-red regularisation for field theory on the Moyal plane, but care is required when taking advantage of this limit because of the global differences between the sphere and the plane. In this section we show that the solitons constructed in section~\ref{largetheta} do converge in a weak sense to solitons on the Moyal plane.
The relevant limit is to send $j$ and $R$ to $\infty$, keeping the ratio
\beq
\theta={ R^2\over 2j}
\eeq
fixed. In theorem 1 of section \ref{largetheta}, we found diagonal solitons for $\theta$ large enough. For each $0<N\leq 2j$, we constructed a soliton $\psi_k^{(j)}$ with the first $N$ eigenvalues lying in the interval $(v,s)$ and with the others in $(0,u)$ (see figure \ref{fig1}). The constraint on $\theta$ is independent of $j$. We will show in theorem 4 below that the solutions $\psi_k^{(j)}$ have the property that there exist $\psi_k$ with
\beq
\psi_k^{(j)} \rightarrow \psi_k \hbox{ as }j \rightarrow \infty
\eeq
at fixed $k$. Assuming this result for the moment, we can take the limits of equation (\ref{shiftelts1}) at fixed $q$ and discover
\beq
\label{moyal1}
\psi_{q+1}-\psi_q = {\theta \over 2(q+1)} \sum_{k=0}^q V^\prime (\psi_k)\, , \;\;\; q=0,1,\cdots
\eeq
which is precisely the equation for a diagonal soliton on the Moyal plane with noncommutativity parameter $\theta$ \cite{Gopakumar:2000zd,Durhuus:2000uz}. We will in addition see that $\psi_k \rightarrow 0$ as $k \rightarrow \infty$ which is necessary for the Moyal plane soliton to have finite energy.\footnote{We note here that the stability analysis for the fuzzy sphere does not carry over to the Moyal plane by simply taking this limit. The condition given on $R$ for stability does not match the condition on $\theta$ for the planar limit, and the more technically complicated analysis given in \cite{Durhuus:2001nj} is therefore still required for the Moyal plane.}\\

\noindent
\begin{bf}{Theorem 4}\end{bf}
{\it Send $j \rightarrow \infty$ and $R \rightarrow \infty$ keeping $\theta$ fixed and sufficiently large. In this limit, the matrix elements of the fuzzy sphere solitons constructed in section \ref{largetheta} converge to the matrix elements of a soliton on the Moyal plane.}
\\

\noindent
We begin by showing that $\psi_0^{(j)}=\inf A^{(j)}$ is a decreasing sequence and so has a limit. For each $j$, we define the interval $I_j = [x^{(j)},y^{(j)}]$, where $x^{(j)},y^{(j)}$ are the points defined in the proof of theorem 1,  so that $A^{(j)} \subset I_j$. We recall that $x^{(j+1)} \leq x^{(j)}$ and $y^{(j+1)} \leq y^{(j)}$. Therefore, if $I_j \cap I_{j+1}= \emptyset$, we have $\inf A^{(j+1)} \leq \inf A^{(j)}$ and we are done. 

If $I_j \cap I_{j+1} \neq \emptyset$, then pick $\alpha \in I_j \cap I_{j+1}$ and suppose $\alpha \notin A^{(j+1)}$. If we set $\psi_0^{(j)}=\psi_0^{(j+1)}=\alpha$, then $\alpha \notin A^{(j+1)}$ means that 
\beq\label{ind0}
\sum_{k=0}^q V'(\psi_k^{(j+1)}) \leq 0 \, , \;\;\; \forall q.
\eeq
We claim that
\beq
\label{ind}
\psi_k^{(j)} \leq \psi_k^{(j+1)}\, , \;\;\; \forall k.
\eeq
By assumption, this is true for $k=0$. We assume that (\ref{ind}) is true for $k \leq n$ and argue by induction. Since, for fixed $k$, $\psi_k^{(j+1)}$ and $\psi_k^{(j)}$ both lie in a domain in which $V'$ is increasing, we have
\beq\label{ind2}
V'(\psi_k^{(j)}) \leq V'(\psi_k^{(j+1)})
\eeq 
for $k \leq n$. Then
\bea
\psi_{n+1}^{(j+1)} - \psi_{n}^{(j+1)}&=& {\theta \over 2(n+1)(1-{n \over 2(j+1)})} \sum_{k=0}^n
V'(\psi_{k}^{(j+1)}) \nonumber \\
& \geq & {\theta \over 2(n+1)(1-{n \over 2(j+1)})} \sum_{k=0}^n
V'(\psi_{k}^{(j)}) \nonumber \\
& \geq & {\theta \over 2(n+1)(1-{n \over 2j})} \sum_{k=0}^n
V'(\psi_{k}^{(j)})
\eea
and, by the induction hypothesis, this implies that
\bea
\psi_{n+1}^{(j+1)} &\geq& \psi_{n}^{(j)} +{\theta \over 2(n+1)(1-{n \over 2j})} \sum_{k=0}^n
V'(\psi_{k}^{(j)}) \nonumber \\ &=& \psi_{n+1}^{(j)},
\eea
establishing (\ref{ind}). Next, using (\ref{ind0}) and (\ref{ind2}), we have
\beq
\sum_{k=0}^q V'(\psi_k^{(j)}) \leq 0 \;\;\; \forall q.
\eeq
which means $\alpha \notin A^{(j)}$.

So far, we have shown that if $\alpha \notin A^{(j+1)}$ then $\alpha \notin A^{(j)}$ and this implies
\beq
I_{j+1} \cap A^{(j)} \subset I_j \cap A^{(j+1)} \subset A^{(j+1)}.
\eeq
Taking the infimum,
\beq
\inf (I_{j+1} \cap A^{(j)}) = \inf A^{(j)} \geq \inf A^{(j+1)},
\eeq
so $\psi_0^{(j)}= \inf A^{(j)}$ is a bounded, decreasing sequence, and so has a limit, $\psi_0^{(j)} \rightarrow \psi_0$, as $j \rightarrow \infty$.

The next stage in proving theorem 4 is to show that, for each fixed $k$, there is a $\psi_k$ such that $\psi_k^{(j)} \rightarrow \psi_k$ as $j \rightarrow \infty$. This follows easily by induction. Assume it is true for $ k\leq n$ and use (\ref{shiftelts1}) to note
\bea
\psi_{n+1}^{(j)}&=&\psi_{n}^{(j)} + {\theta \over 2(n+1)(1-{n \over 2j})} \sum_{k=0}^n V' (\psi_k^{(j)})\\
& \rightarrow &\psi_{n} + {\theta \over 2(n+1)} \sum_{k=0}^n V' (\psi_k)
\eea
as $j \rightarrow \infty$ by continuity of $V'$.

Finally, it remains to check the finite energy condition $\psi_k \rightarrow 0$ as $k \rightarrow \infty$. We first note that, since any soliton on the fuzzy sphere has its spectrum in the interval $[0,s]$, we have $\psi_k^{(j)}\geq 0$ for all $k,j$, and this implies $\psi_k \geq 0$ for all $k$. Similarly, since $\psi_{k+1}^{(j)} \leq \psi_{k}^{(j)}$, we also have $\psi_{k+1} \leq \psi_{k}$. Then $\psi_{k+1}$ is a decreasing sequence which is bounded below and so there exists $\psi \geq 0$ with  $\psi_{k} \rightarrow \psi$ as $k \rightarrow \infty$.

By our original construction, we also have $\psi_k^{(j)}\leq u$ for all $j$ and $k >N$, and so, taking the limit, $\psi_k \leq u$ for $k>N$. Then only $\psi_0, \cdots , \psi_N$ lie outside the region in which $V'$ is increasing, so $V'(\psi_k)>V'(\psi)$ for $k \geq N+1$. Substituting into (\ref{moyal1}) gives
\beq
\psi_{q+1}-\psi_{q} > {\theta q \over 2(q+1)} V'(\psi) + {\theta \over 2(q+1)} \sum_{k=0}^{N} V'(\psi_k)
\eeq
for $q>N$, and taking the limit
\beq
0 \geq {\theta  \over 2}V'(\psi) \;\;\;  (\geq 0)
\eeq
and we deduce that $\psi=0$. So $\psi_k \rightarrow 0$ as $k \rightarrow \infty$, and this completes the proof of theorem~4.

\acknowledgments{We are grateful to Robert Magnus for valuable discussions.\ 
This research was partly supported by TMR grant no. HPRN-CT-1999-00161, and grants from the Icelandic Research Council, and the University of Iceland Research Fund.}

\providecommand{\href}[2]{#2}\begingroup\raggedright\endgroup

\end{document}